\documentclass[aps,prb,twocolumn,superscriptaddress]{revtex4}
\usepackage{amsfonts}
\usepackage{amssymb}
\usepackage{amsmath}
\usepackage{graphicx}
\usepackage{epsfig}
\usepackage{color}
\usepackage{amsmath,bm}

\begin{document}

\title{$\eta $-pairing ground states in the non-Hermitian Hubbard model}
\author{X. Z. Zhang}
\affiliation{College of Physics and Materials Science, Tianjin Normal University, Tianjin
300387, China}
\author{Z. Song}
\email{songtc@nankai.edu.cn}
\affiliation{School of Physics, Nankai University, Tianjin 300071, China}

\begin{abstract}
The introduction of non-Hermiticity has greatly enriched the research field
of traditional condensed matter physics, and eventually led to a series of
discoveries of exotic phenomena. We investigate the effect of non-Hermitian
imaginary hoppings on the attractive Hubbard model. The exact bound-pair
solution shows that the electron-electron correlation suppresses the
non-Hermiticity, resulting in off-diagonal long-range order (ODLRO) ground
state. In a large negative $U $ limit, the ODLRO ground state corresponds to
$\eta $-spin ferromagnetic states. We also study the system with mixed
hopping configuration. The numerical result indicates the existence of the
transition from normal to $\eta $-pairing ground states by increasing the
imaginary hopping strength. Our results provide a promising approach for the
non-Hermitian strongly correlated system.
\end{abstract}

\maketitle

\section{Introduction}

\label{introduction} Non-Hermitian systems that can only be described by
non-Hermitian Hamiltonians are ubiquitous in nature. Many open systems,
which are not fully isolated from the rest of world, belong to this class.
Comparing to the Hermitian systems, the probability of the non-Hermitian
system effectively becomes nonconserving due to the exchange of energy,
particles, and information with external degrees of freedom that are out of
the Hilbert space. Mainly driven by experimental progress in atomic physics
\cite{Bloch2008,DAlessio2016}, the last two decades have witnessed
remarkable developments in studies of out-of-equilibrium dynamics in
isolated quantum many-body systems. It has become possible to study open
many-body physics in a highly controlled manner \cite%
{Syassen2008,Barontini2013,Patil2015,Rauer2016,Lueschen2017,Li2019a,Corman2019}%
. Within this burgeoning field, the treasure hunt is sprouting into
fascinating new directions ranging from non-Hermitian extensions of Kondo
effect \cite{Nakagawa2018,Lourenfmmodeboxclsecio2018}, many-body
localization \cite{Hamazaki2019}, to fermionic superfluidity \cite%
{Yamamoto2019,Okuma2019}.

Recent advances in quantum simulations of the Hubbard model with ultracold
atoms have offered a multifunctional platform to unveil low-temperature
properties of the strongly correlated system \cite%
{Bakr2009,Parsons2015,Cheuk2015,Cheuk2016,Parsons2016,Esslinger2020}. A
series of cornerstone works have reshaped our understanding of the
dissipative strongly correlated system \cite%
{Syassen2008,Fausti2011,Hu2014,Zhu2014,Kaiser2014,Mitrano2016,Tomita2017,Sponselee2018,Cantaluppi2018,Sato2019,McIver2020,Booker2020,Tindall2020,Zhang2020a}%
. One of the most tantalizing findings is the possible superconductivity in
which $\eta $-pairing state plays a vital role. This stimulates a plethora
of non-equilibrium protocols including photodoping schemes \cite%
{Iwai2003,Rosch2008,Sensarma2010,Eckstein2011,Ichikawa2011,Lenarfmmodeheckclseciifmmodeheckclseci2013,Stojchevska2014,Mitrano2014,Werner2019a,Peronaci2020,Li2020}
and dissipation-induced schemes \cite%
{Diehl2008,Kraus2008,Coulthard2017,Werner2019,Zhang2020b} to selectively
generate such superconducting-like states. However, few people discuss the
impact of non-Hermiticity on the low-lying energy spectrum and quantum
magnetism of the strongly correlated system from the level of non-Hermitian
quantum mechanics \cite{Nakagawa2020}.

It is the aim of this paper to investigate the effect of non-Hermiticity on
the strongly correlated system in the context of the non-Hermitian quantum
mechanics. We show that the non-Hermitian imaginary hopping can indeed
induce a robust $\eta $-pairing ground state for a wide range of parameters $%
U $ (particle-particle interaction) and $t$ (hopping strength), by
considering the bipartite non-Hermitian Hubbard system. An exact solution of
the bound pair is employed to elucidate the underlying paring mechanism and
pave the way to extend the results to dilute gas. In physics, the
particle-particle interaction suppresses the non-Hermiticity leading to the
off-diagonal long-range order (ODLRO) ground state with real energy, and the
non-Hermitian imaginary hopping, in turn, suppresses the antiferromagnetic
correlation, thus ensuring the system has a $\eta $-spin ferromagnetic
ground state. Numerical results of 1D and 2D systems with corrugation
patterns indicate that such property is insensitive to the disorder and the
strength of the interaction even though the on-site interaction breaks the $%
SO(4)$ symmetry, which suggests a promising scheme in a real experiment. We
further demonstrate that there can exist a transition from normal to $\eta $%
-pairing ground state associating with the sudden change of the
doublon-doublon correlation.

The paper is organized as follows. Sec. ~\ref{model} discusses the
non-Hermitian Hubbard model, the non-Hermiticity of which originates from
the imaginary hopping. Sec. ~\ref{bp} introduces the exactly two-particle
solution providing the mechanism of the formation of the $\eta $-pairing
ground state. Sec. ~\ref{largeU} gives the effective magnetic Hamiltonian
under the large negative $U$ limit. Sec. ~\ref{mixed} shows the numerical
results and the analytical understanding of superconductive $\eta $-pairing
ground state. Sec. ~\ref{transition} demonstrates the transition from a
normal to superconductive ground state. Sec. ~\ref{summary} concludes this
paper. Some details of our calculations are placed in the Appendixes.

\section{Model}

\label{model} We consider a non-Hermitian Hubbard model on a bipartite
lattice%
\begin{eqnarray}
H &=&i\sum_{j,l}\sum_{\sigma =\uparrow ,\downarrow }t_{jl}(c_{j,\sigma
}^{\dagger }c_{l,\sigma }+c_{l,\sigma }^{\dagger }c_{j,\sigma })  \notag \\
&&+U\sum_{j}n_{j,\uparrow }n_{j,\downarrow },  \label{non_H}
\end{eqnarray}%
with the following notation: the operator $c_{j,\sigma }$ ($c_{j,\sigma
}^{\dagger }$) is the usual annihilation (creation) operator of a fermion
with spin $\sigma \in \left\{ \uparrow ,\downarrow \right\} $ at site $j$,
and $n_{j,\sigma }=c_{j,\sigma }^{\dagger }c_{j,\sigma }$ is the number
operator for a particle of spin $\sigma $ on site $j$; the symbol $i=\sqrt{-1%
}$ represents an imaginary number; $U$ and $t_{jl}$ are required to be real
and play the role of interaction and kinetic energy scales, respectively;
the system can be divided into two sublattices $A$ and $B$ such that $%
t_{jl}=0$ whenever $j\in \left\{ A\right\} $ and $l\in \left\{ A\right\} $
or $j\in \left\{ B\right\} $ and $l\in \left\{ B\right\} $. The
non-Hermiticity of $H$ stems from the imaginary hopping $it_{jl}$ that can
be realized by the judicious design of the loss and the magnetic flux \cite%
{Li2015,Li2017} which are within the reach of cold atom experiments \cite%
{Lee2014,Schreiber2015}. Notice that such non-Hermiticity is distinct from
the complex particle-particle interaction adopted to describe the inelastic
collision of two particles \cite{Zhang2017,Nakagawa2020}. When the uniform
Hermitian hopping is taken, the Hamiltonian can feature a Mott insulating
ground state with a strong antiferromagnetic correlation that is generically
nonsuperconducting. Evidently, the imaginary hopping inevitably competes
with the interaction leading to the unique properties of the considered
system. It can be expected that the introducing of such non-Hermiticity will
significantly alter the magnetic correlation of the system.

In this paper, we focus on whether the system can favor the ground state
with superconductivity in this non-Hermitian setting. To gain physical
insight into this system, we first investigate the symmetry of the
considered model. It has two sets of commuting $SU(2)$ symmetries. The first
is the spin symmetry characterized by the generators%
\begin{eqnarray}
s^{+} &=&\left( s^{-}\right) ^{\dagger }=\sum_{j}s_{j}^{+}, \\
s^{z} &=&\sum_{j}s_{j}^{z},
\end{eqnarray}%
where the local operators $s_{j}^{+}=c_{j,\uparrow }^{\dagger
}c_{j,\downarrow }$ and $s_{j}^{z}=\left( n_{j,\uparrow }-n_{j,\downarrow
}\right) /2$ obey the Lie algebra, i.e., $[s_{j}^{+},$ $%
s_{j}^{-}]=2s_{j}^{z} $, and $[s_{j}^{z},$ $s_{j}^{\pm }]=\pm s_{j}^{\pm }$.
Large values of the spin quantum number $s$ corresponds to ferromagnetism.
The second often referred to as $\eta $ symmetry has the generators%
\begin{eqnarray}
\eta ^{+} &=&\left( \eta ^{-}\right) ^{\dagger }=\sum_{j}\eta _{j}^{+}, \\
\eta ^{z} &=&\sum_{j}\eta _{j}^{z},
\end{eqnarray}%
with $\eta _{j}^{+}=\lambda c_{j,\uparrow }^{\dagger }c_{j,\downarrow }$ and
$\eta _{j}^{z}=\left( n_{j,\uparrow }+n_{j,\downarrow }-1\right) /2$
satisfying commutation relation, i.e., $[\eta _{j}^{+},$ $\eta
_{j}^{-}]=2\eta _{j}^{z}$, and $[\eta _{j}^{z},$ $\eta _{j}^{\pm }]=\pm \eta
_{j}^{\pm }$. Here we assume a bipartite lattice and $\lambda =1$ for $j\in
\left\{ A\right\} $ and $-1$ for $j\in \left\{ B\right\} $. Notice that
under a particle-hole transformation, $c_{j,\downarrow }\rightarrow \lambda
c_{j,\downarrow }^{\dagger }$, which maps the attractive Hubbard model to a
repulsive one in the parent Hermitian Hamiltonian (\ref{non_H}), the role of
the two sets of $SU(2)$ generators is interchanged. Straightforward algebra
shows that
\begin{eqnarray}
\left[ H,\eta ^{\pm }\right] &=&\pm U\eta ^{\pm }, \\
\left[ H,\eta ^{z}\right] &=&0,
\end{eqnarray}%
which indicates that one can construct many exact eigenstates $H\left( \eta
^{+}\right) ^{N}\left\vert \mathrm{Vac}\right\rangle =NU\left( \eta
^{+}\right) ^{N}\left\vert \mathrm{Vac}\right\rangle $ with $\left\vert
\mathrm{Vac}\right\rangle $ being the vacuum state of fermion $c_{j,\sigma }$%
. Correspondingly, the large values of the $\eta $ quantum number are
related to a staggered ODLRO and superconductivity \cite{Yang1962,Singh1991}%
.
\begin{figure*}[tbp]
\centering
\includegraphics[width=0.7\textwidth]{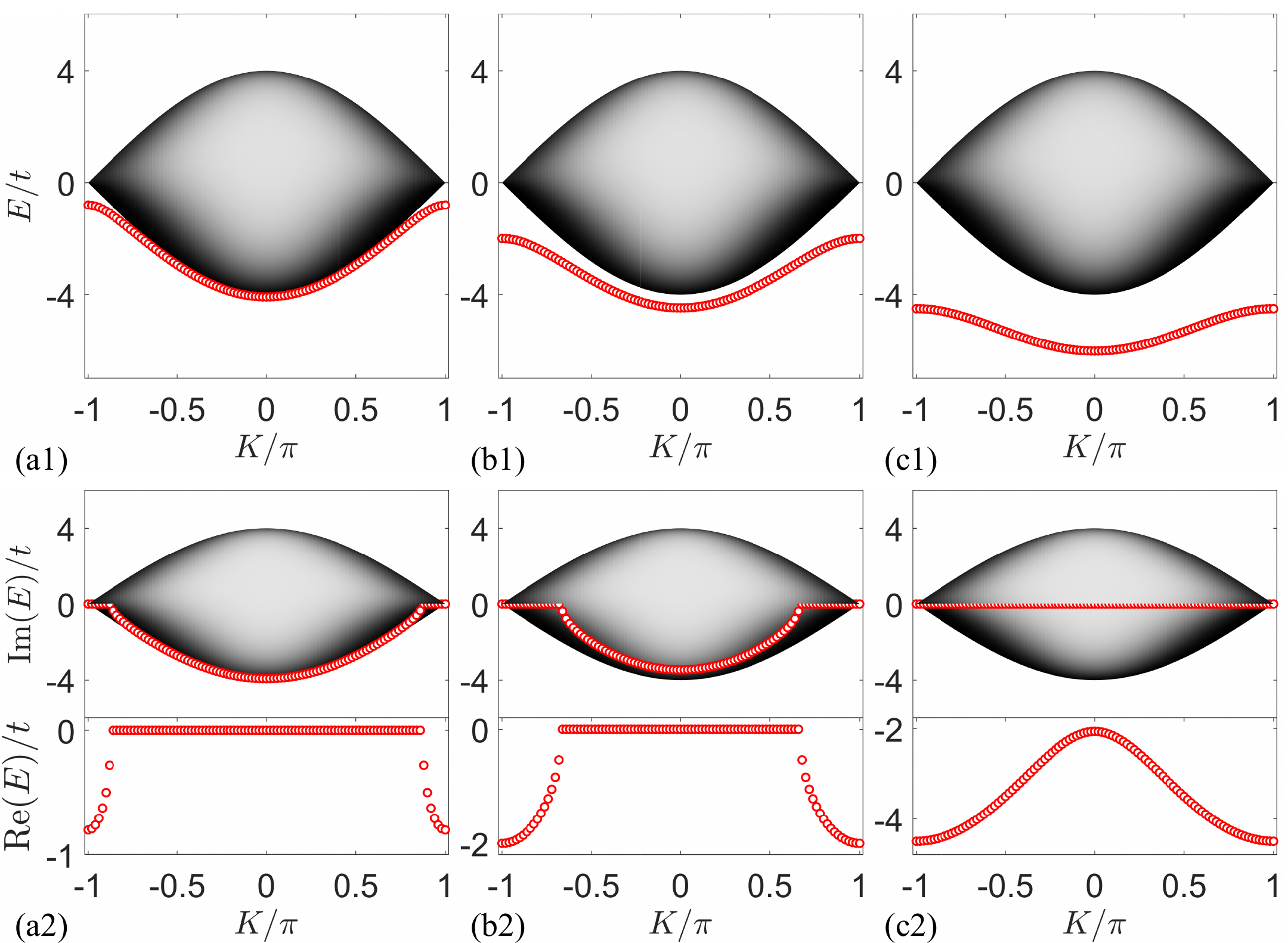}
\caption{Comparison of the two-particle spectrum within the subspace $(0,0)$
between the non-Hermitian setting and its parent Hermitian system for (a) $%
U=-0.8t$, (b) $U=-2t$, and (c) $U=-4.5t$, respectively. The upper and lower
panels present the spectrum of the Hermitian and non-Hermitian system,
respectively. The red circle and gray shading denote the bound pair and
scattering state. The parent Hermitian system can be obtained by assuming $%
it\rightarrow t$. For the Hermitian system, the bound pair with the lowest
energy lies in the $K=0$ subspace while the ground state of two-particle
non-Hermitian setting locates on the subspace indexed by $K=\protect\pi $.
It is shown that the presence of the imaginary hopping not only makes all
scattering energy bands imaginary but also reverses the whole bound band. In
the condition of small $U$, there can exist an EP characterized by the
divergence of $\partial \protect\epsilon _{K}/\partial K$. Such
non-Hermiticity alters significantly the paring mechanism and hence favors
superconductivity.}
\label{fig_spectrum}
\end{figure*}
\begin{figure*}[tbp]
\centering
\includegraphics[width=0.7\textwidth]{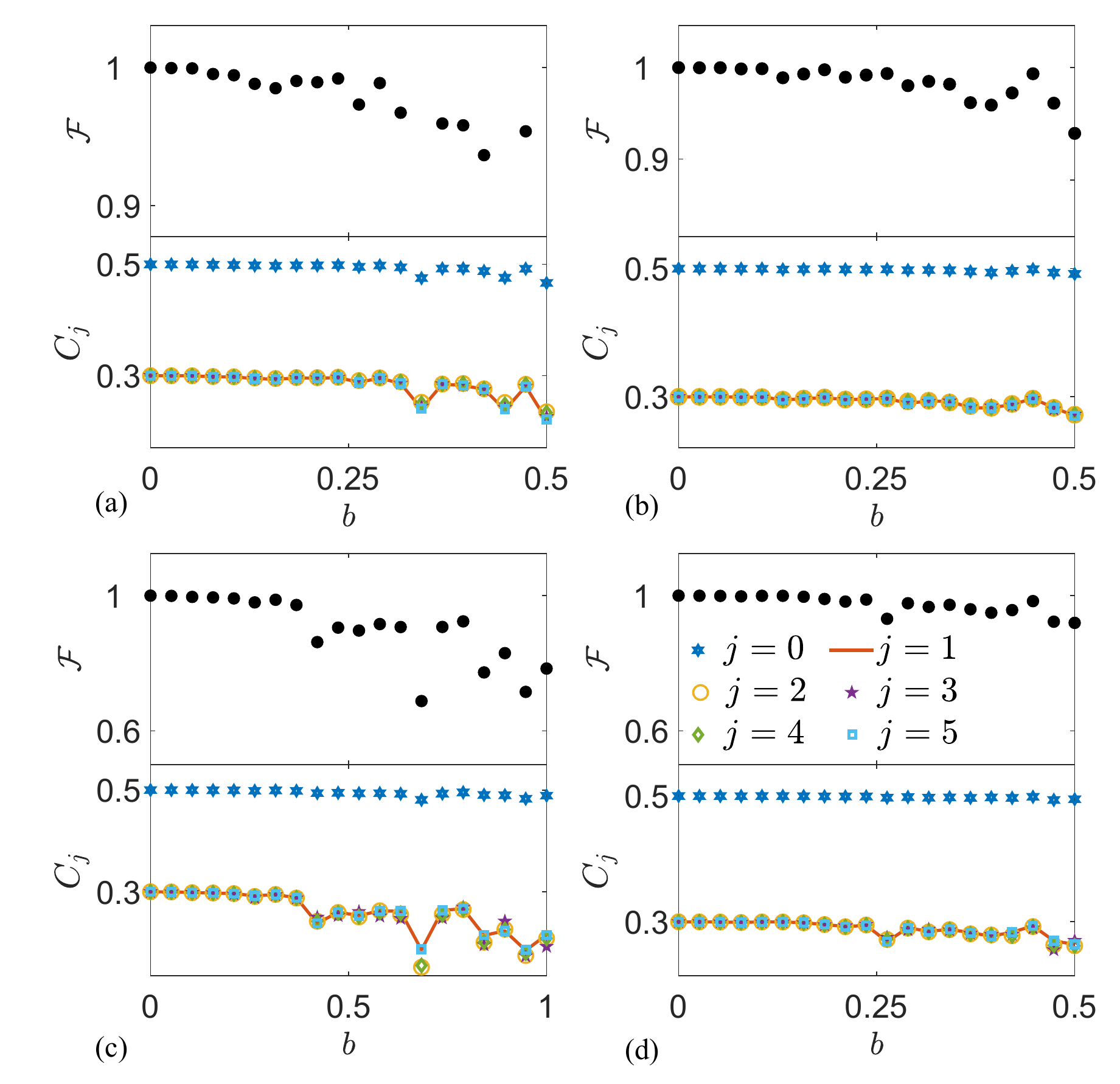}
\caption{Plots of the overlap $\mathcal{F}$ and correlator $C_{j}$ as a
function of the strength of interaction disorder $b$ for (a) $t=1$, $a=0.1t$%
, $U=-0.5t$ (b) $t=1$, $a=0.3t$, $U=-1.5t$ (c) $t=1$, $a=0$, $U=-4t$, and
(d) $t=1$, $a=0.2t$, $U=-4t$. The numerical simulation is performed for the $%
6$ site 1D Hubbard model at half filling and $s^{z}=0$. Here the strength of
the hopping disorder $a$ is set to be constant for each subfigure and the
correlator $C_{j}$ is averaged over all sites separated by a distance $j$.
When $b=0$, no matter what value $a$ takes, as long as $U$ is non-zero, one
can always get a perfect $\protect\eta $-pairing ground state. The variation
of $C_{j}$ indicates that the increase of $b$ will not result in the
significant change of the $\protect\eta $-pairing ground state; the presence
of the hopping disorder can suppress the fluctuation of $C_{j}$ compared to
the disorder free case, which can be seen from (c)-(d). Therefore, the value
of the correlator is the consequence of the interplay between two such
disorders, which provides a scheme to prepare $\protect\eta $-pairing ground
state in the experiment.}
\label{fig_ran}
\end{figure*}

\section{$\protect\eta $-pairing state in two-particle subspace}

\label{bp} Based on the symmetry of the system, we first elucidate the
paring mechanism through the exact solution within the two-particle
subspace. Supposing that the Hamiltonian (\ref{non_H}) describes a 1D
homogeneous ring system in which $it_{jl}=it$. Owing to the translation
symmetry, the basis of such invariant subspace can be constructed as follow%
\begin{eqnarray}
\left\vert \phi _{0}^{-}\left( K\right) \right\rangle &=&\frac{1}{\sqrt{N}}%
\sum_{j}e^{iKj}c_{j,\uparrow }^{\dagger }c_{j,\downarrow }^{\dagger
}\left\vert \mathrm{Vac}\right\rangle , \\
\left\vert \phi _{r}^{\pm }\left( K\right) \right\rangle &=&\frac{1}{\sqrt{2N%
}}e^{iKr/2}\sum_{j}e^{iKj}(c_{j,\uparrow }^{\dagger }c_{j+r,\downarrow
}^{\dagger } \\
&&\pm c_{j,\downarrow }^{\dagger }c_{j+r,\uparrow }^{\dagger })\left\vert
\mathrm{Vac}\right\rangle ,
\end{eqnarray}%
and%
\begin{equation}
\frac{s^{\pm }}{\sqrt{2}}\left\vert \phi _{r}^{+}\left( K\right)
\right\rangle =\frac{1}{\sqrt{N}}e^{iKr/2}\sum_{j}e^{iKj}c_{j,\pm \uparrow
}^{\dagger }c_{j+r,\pm \uparrow }^{\dagger }\left\vert \mathrm{Vac}%
\right\rangle ,
\end{equation}%
where $N$ is an even number and $K=2n\pi /N$ is the momentum vector indexing
the subspace. $r$ represents the relative distance between the two
particles. These bases are eigenvectors of the operators $s^{2}$ and $s^{z}$%
, which satisfies
\begin{eqnarray}
s^{2}\left\vert \phi _{r}^{-}\left( K\right) \right\rangle &=&0,\text{ } \\
s^{z}\left\vert \phi _{r}^{-}\left( K\right) \right\rangle &=&0, \\
s^{2}\left\vert \phi _{r}^{+}\left( K\right) \right\rangle &=&2\left\vert
\phi _{r}^{+}\left( K\right) \right\rangle ,\text{ } \\
s^{z}\left\vert \phi _{r}^{+}\left( K\right) \right\rangle &=&\left\vert
\phi _{r}^{+}\left( K\right) \right\rangle .
\end{eqnarray}%
Evidently, each subspace labeled by $K$ can be further decomposed into four
subspaces with $(s,$ $s^{z})=(0,$ $0)$, $(1,$ $0)$ and $(1,$ $\pm 1)$ in
term of spin symmetry. Aiding by the detailed calculation in the Appendix,
the bound pair emerges in the $(0,$ $0)$ subspace with eigen energy being $%
\epsilon _{K}=\mathrm{sgn}\left( U\right) \sqrt{U^{2}+4\lambda _{K}^{2}}$ in
which $\lambda _{K}=2it\cos \left( K/2\right) $. The bound pair state is $%
\left\vert \varphi _{K}^{\mathrm{b}}\right\rangle =\sum_{r}f_{K}^{-}\left(
r\right) \left\vert \phi _{r}^{-}\left( K\right) \right\rangle $ with
\begin{equation}
f_{K}^{-}\left( j\right) =\left\{
\begin{array}{c}
1/\sqrt{2}\text{, }j=0 \\
e^{-\beta j}\text{, }j\neq 0%
\end{array}%
\right. ,
\end{equation}%
where $\beta =\ln [(-U\pm \sqrt{U^{2}+4\lambda _{K}^{2}})/2\lambda _{K}]$.
Here $\pm $ denotes negative and positive $U$, respectively. We concentrate
on the negative $U$ in the following unless stated otherwise. In the absence
of on-site interaction $U$, only the scattering eigenstate with imaginary
eigenenergy presents and the system does not accommodate the bound pair
state. The nonzero interaction $U$ leads to the emergence of the bound pair.
When $\left\vert U\right\vert >\left\vert 4t\right\vert $, the system
possesses the full real bound pair spectrum. However, a small $U$ results in
the appearance of the imaginary bound pair energy. The corresponding
eigenstate is in the form of an oscillation damping wave rather than a
monotonic damping wave of the Hermitian parent system. Notice that if $%
\left\vert U\right\vert $ $\leqslant $ $\left\vert 4t\right\vert $, then an
exceptional point (EP) $\left\vert U\right\vert =\left\vert 2\lambda
_{K_{c}}\right\vert $ presents, at which the coalescent eigenstate
approaches to a unidirectional plane wave with $\beta =0$ or $\pi $
corresponding to $K=0$ or $2\pi $. In this sense, the non-Hermiticity of the
system is suppressed through the pairing mechanism. The emergence of real
energy is the consequence of the competition between the on-site interaction
and imaginary hopping. Furthermore, the lowest real eigenenergy appears in
the $K=\pi $ subspace no matter whether the system possesses the full real
spectrum. The corresponding ground state is $\eta $-pairing state with the
form
\begin{equation}
\left\vert \phi _{0}^{-}\left( K\right) \right\rangle =\left( \eta
^{+}\right) /\sqrt{N}|\mathrm{Vac}\rangle ,
\end{equation}
and thus it favors superconductivity. This is in stark difference from the
Hermitian system, i.e., $it\rightarrow t$. In that case, the ground state of
the two-particle system locates on the $K=0$ rather than $K=\pi $ subspace
such that the $\eta $-pairing state has the highest eigenenergy than the
other bound pair state. Fig. \ref{fig_spectrum} shows the typical energy
spectrum of subspace $(0,0)$. It demonstrates that the imaginary hopping
flips the bound pair spectrum of the parent Hermitian spectrum so that $\eta$%
-pairing state becomes the ground state of the system. It is worthy pointing
out that if we consider the dilute Fermi gas formed by many bound pairs in
which the pair-pair interaction is neglected, then the mechanism for a
single bound pair can be extended to this type of dilute gas.
\begin{figure*}[tbp]
\centering
\includegraphics[width=0.8\textwidth]{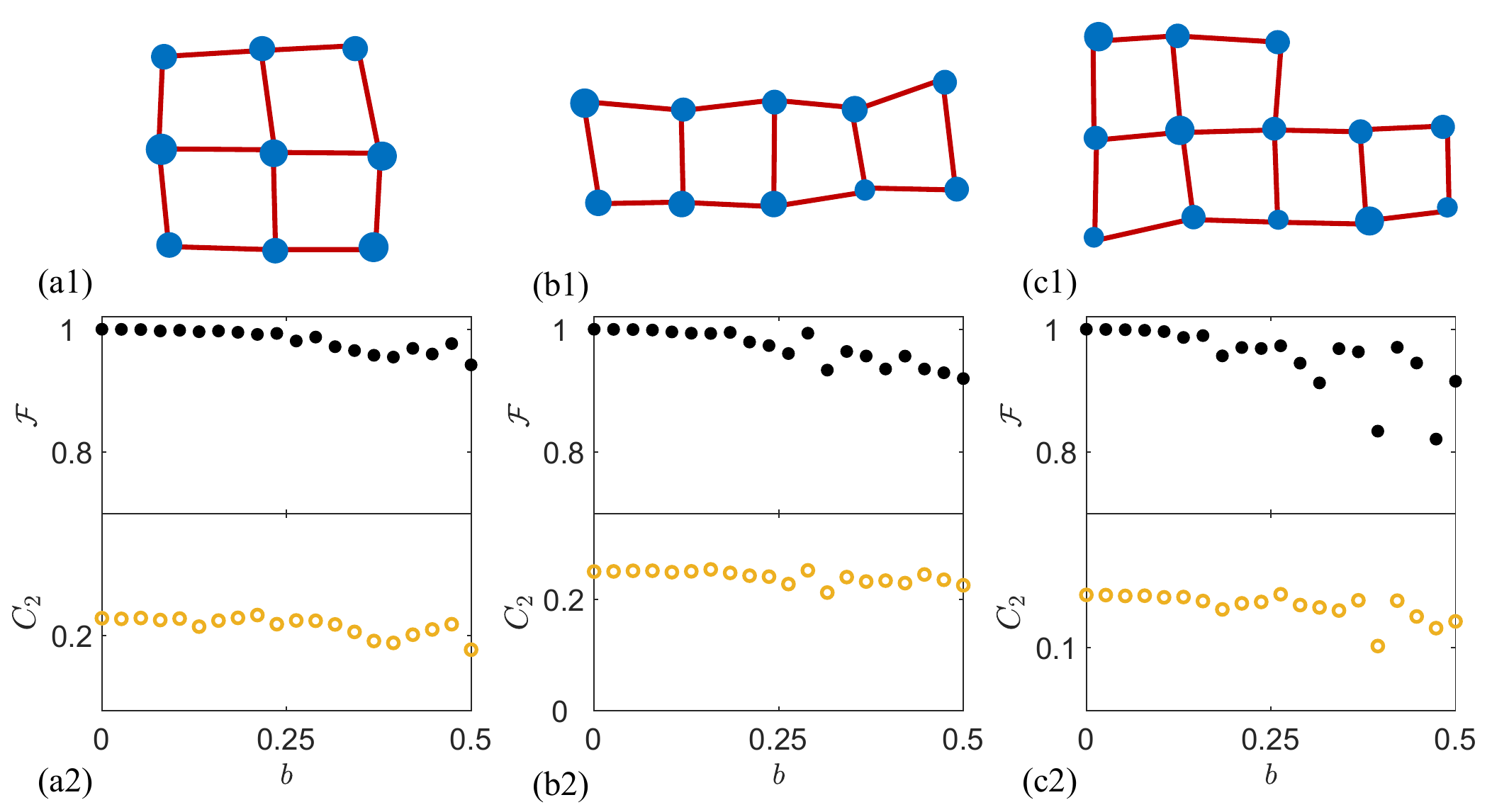}
\caption{Numerical simulation of 2D corrugation pattern for (a1) $8$ site
Hubbard model with $4$ filled particles, (b1) $9$ site Hubbard model with $6$
filled particles, and (c1) $13$ site Hubbard model with $2$ filled
particles. The different sizes of solid circles and different lengths of red
edges denote the disorder of the interaction strength $U_{j}$ and hopping
strength $t_{j}$ of Eq. (\protect\ref{disorder}). (a2)-(c2) Plots of $%
\mathcal{F}$ and $C_{1}$ as function of disorder strength $b$. The system
parameters are (a) $t=1$, $a=0$, $U=-0.8t$ (b) $t=1$, $a=0.2t$, $U=-0.8t$
and (c) $t=1$, $a=0.2t$, $U=-0.5t$. Similar with 1D system, the induced
fluctuation of $C_{1}$ is minor in comparison with disorder free case ($b=0$%
) indicating that the existing result of 1D can be extended to 2D or higher
dimensional system.}
\label{fig_2D}
\end{figure*}

\section{$\protect\eta $-pairing state in the large $U$ limit}

\label{largeU} Now we turn to investigate the situation with arbitrary
filling but in the large $U$ limit. system. Following the standard step of
quantum mechanics, the system can be divided into the kinetic part $%
H^{\prime }$ and interaction part $H_{0}$, where
\begin{eqnarray}
H^{\prime } &=&i\sum_{j,l}\sum_{\sigma =\uparrow ,\downarrow
}t_{jl}(c_{j,\sigma }^{\dagger }c_{l,\sigma }+c_{l,\sigma }^{\dagger
}c_{j,\sigma }), \\
H_{0} &=&U\sum_{j}n_{j,\uparrow }n_{j,\downarrow }.
\end{eqnarray}%
Here the imaginary hopping is assumed to be homogeneous $it_{jl}=it$. In the
strongly correlated regime $\left\vert U\right\vert \gg t$, the kinetic term
$H^{\prime }$ can be treated as a perturbation and one can derive an
effective Hamiltonian for the degenerate space. To second order in
perturbation theory, the effective Hamiltonian is given by%
\begin{equation}
H_{\mathrm{eff}}=P_{0}H_{0}P_{0}+P_{0}H^{\prime }P_{1}\frac{1}{E_{0}-H_{0}}%
P_{1}H^{\prime }P_{0}+O\left( \frac{t^{3}}{U^{2}}\right) ,
\end{equation}%
where $P_{0}$ is a projector onto the Hilbert subspace in which there are $M$
lattice sites occupied by two particles with opposite spin orientation, and $%
P_{1}=1-P_{0}$ is the complementary projection. Here the energy $E_{0}$ of
the unperturbed state is set to $E_{0}=MU$ where $M$ denotes the number of
doublons. Since $H^{\prime }$ acting on states in $P_{0}$ annihilates only
one double occupied site, all states in $P_{1}H^{\prime }P_{0}$ have exactly
$N-1$ doubly occupied sites. Therefore, the effective Hamiltonian regarding
doublon-hole creation and recombination process is given as%
\begin{equation}
H_{\mathrm{eff}}=MU+\frac{4t^{2}}{U}\sum_{j}\left( \bm{\eta }_{j}\cdot %
\bm{\eta }_{j+1}-\frac{1}{4}\right) ,  \label{heff}
\end{equation}%
where $\bm{\eta }_{j}=(\eta _{j}^{x},$ $\eta _{j}^{y},$ $\eta _{j}^{z})$. In
the Appendix, a simple two-site case is provided to elucidate this
mechanism. This indicates that the non-Hermitian virtual exchange mediates
an interaction between the pseudo spins. It is similar to the Heisenberg
interaction in the Hermitian Hubbard model in the way that the doubly
occupied and vacuum sites correspond to spin up and spin down states,
respectively. Owing to the fact that $4t^{2}/U<0$, the effective Hamiltonian
is ferromagnetic Heisenberg model of pseudo spins. The eigenstate of the
lowest eigenenergy within each doublon subspace is the $\eta $-pairing state
with different pair number. As such the ground state of $H_{\mathrm{eff}}$
is $\left( \eta ^{+}\right) ^{N}\left\vert \mathrm{Vac}\right\rangle $
dubbed as $\eta $-spin ferromagnetic state. Note in passing that for the
case of repulsive Hubbard model ($U>0$) at half filling, the extra minus
sign induced by the non-Hermitian virtual exchange leads to an effective
ferromagnetic rather than an antiferromagnetic Heisenberg Hamiltonian
\begin{equation}
H_{\mathrm{eff}}^{\mathrm{l}}=-\frac{4t^{2}}{U}\sum_{j}\left( \bm{s}%
_{j}\cdot \bm{s}_{j+1}-\frac{1}{4}\right)
\end{equation}%
describing the behavior of the ground state and low energy excitations. The
eigenenergy of the ground state is zero. Notably, the interplay between the
imaginary hopping and particle-particle interaction fundamentally alters the
magnetism of the Hubbard model leading to sign reversal of magnetic
correlation.
\begin{figure}[tbp]
\centering
\includegraphics[width=0.35\textwidth]{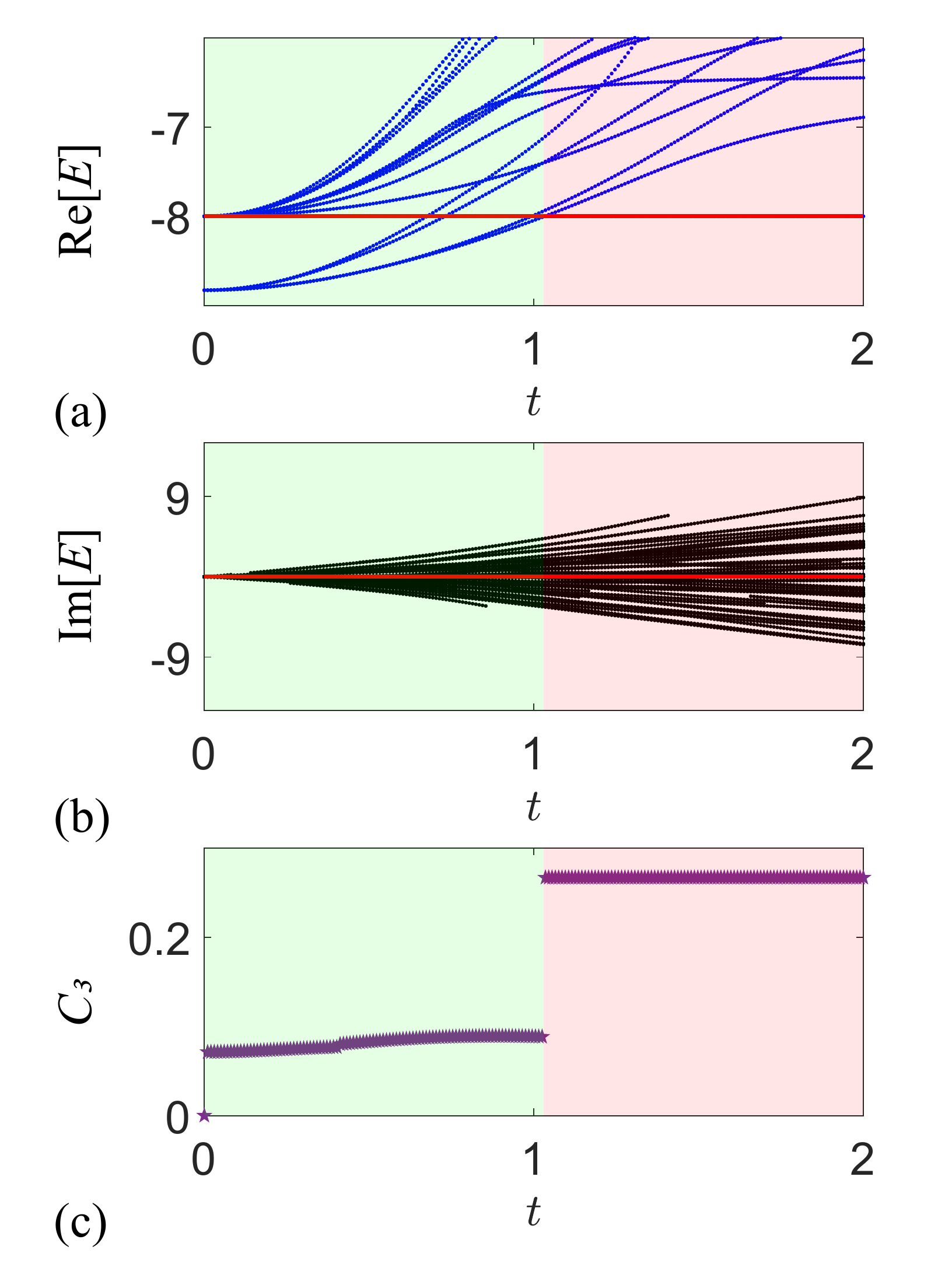}
\caption{Transition of the ground state driven by the imaginary hopping.
(a)-(b) depict the variation of the energy spectrum with respect to hopping
strength $t$, and (c) presents the switch of the correlator $C_{3}$. The
numerical simulation is performed for the $6$ site Hubbard model with $4$
filled particles. The other system parameters are $t_{0}=1$, and $U=-1.5t_{0}
$. The hopping strength $t$ is in units of $t_{0}$. The red line denotes the
$\protect\eta $- pairing state. $C_{3}$ experiences an evident jump at the
critical point indicating a dramatic change of the ground state. Such a
first-order transition may associate with quantum phase transition to some
extent.}
\label{fig_gene}
\end{figure}

\section{$\protect\eta $-pairing state in system with mixed hoppings}

\label{mixed} In the aforementioned sections, we have demonstrated that the $%
\eta $-pairing state can be either the ground state of the system under the
large $U$ limit, or the ground state of the two-particle subspace with
non-zero $U$. Then a natural question arises: (i) For any nonzero $U$, is
the $\eta $-pairing state still the ground state of the system in the
subspace of other particle numbers? (ii) If yes, can the existing 1D results
be extended to 2D or higher dimensional system? (iii) If the disorder is
introduced, does the property of ground state be changed? To answer these
questions, we first investigate the 1D non-Hermitian system with disordered
imaginary hoppings and interaction since the system parameter does not hold
the uniform in real experiments. The corresponding disordered Hamiltonian
can be obtained by taking two sets of random numbers $\left\{ t_{j}\right\} $
and $\left\{ U_{j}\right\} $ around $t$ and $U$ in Eq. (\ref{non_H}). The
random number parameter can be taken as
\begin{equation}
t_{j}=t+\mathrm{rand}(-a,a)\mathrm{\text{, }}U_{j}=U+\mathrm{rand}(-b,b),
\label{disorder}
\end{equation}%
where $\mathrm{rand}(-a,a)$ denotes a uniform random number within $(-a,a)$.
It is too cumbersome to obtain an analytical result. Hence, we perform the
numerical simulation to check the fidelity between the ground state and
target $\eta $-pairing state, which can be given as
\begin{equation}
\mathcal{F}=\left\vert \langle \psi _{g}\left( m\right) \right\vert \psi
_{c}\left( m\right) \rangle |,
\end{equation}%
with $m$ being an even number representing filled particle number of the
system. Here $|\psi _{g}\left( m\right) \rangle $ is the ground state of
such subspace and the target $\eta $-pairing state is
\begin{equation}
|\psi _{c}\left( m\right) \rangle =\Omega ^{-1}\left( \eta ^{+}\right)
^{m/2}|\mathrm{Vac}\rangle ,
\end{equation}%
with renormalization coefficient $\Omega =\sqrt{C_{N}^{m/2}}$. Fig. \ref%
{fig_ran} shows that if $U$ is homogeneous ($b=0$) then the ground state is
the $\eta $-pairing state in the subspace with particle numbers $m=2$, $4$, and $%
6$. The corresponding energies are $U$, $2U$, and $3U$
respectively. Such a result indicates that the formation of the $\eta $%
-pairing ground state does not depend on the values of $U$ and is immune to
the hopping disorder. However, the introduction of disordered $U$ will cause
the ground state of the system to deviate from the $\eta $-pairing state.
The underlying mechanism is clear, that is, the system fulfills the $\eta $
symmetry even in the presence of the hopping disorder, but if one introduce
disorder into the interaction, this symmetry will be destroyed leading to
such deviation. One may think that the disorder of interaction scrambles the
background spin configuration and disturb the spin correlation. Then a
question arises: to what extent does the ground state maintain the
superconductivity? To capture superconductivity, the doublon-doublon
correlator
\begin{equation}
C_{j}=\sum_{i}\langle \eta _{i}^{+}\eta _{i+j}^{-}\rangle /N
\end{equation}%
is introduced. It is averaged over all sites separated by a distance $j$.
The nonzero value of such quantity implies both the Meissner effect and flux
quantization and hence provides a possible definition of superconductivity.
\cite{Yang1962,Yang1989} For the target state $|\psi _{c}\left( m\right)
\rangle $, the expectation value can be given as
\begin{equation}
\langle \psi _{c}\left( m\right) |\eta _{i}^{+}\eta _{i+j}^{-}|\psi
_{c}\left( m\right) \rangle =\left\{
\begin{array}{c}
\frac{M(N-M)}{N(N-1)}\text{, for }j\neq 0 \\
\frac{M}{N}\text{, for }j=0%
\end{array}%
\right. ,
\end{equation}%
where $M=m/2$. Notice that it is irrelative to the distance $j$ and hence
the correlator $C_{j}$ obeys the same law such that $C_{j}=M(N-M)/[N(N-1)]$
for $j\neq 0$ or $C_{j}=M/N$ for $j=0$. Fig. \ref{fig_ran} shows that the
value of correlator and the overlap between the ground state and target
state. It indicates that $\mathcal{F}$ is around $0.9$ and the correlator $%
C_{j}$ stays at a non-zero value ensuring the ground state of the system
possesses the superconductivity even though the strong inhomogeneity of
interaction presents.

Now we switch gears to the cases of the 2D system. In Fig. \ref{fig_2D}, the
disordered 2D system is sketched. For simplicity, we fix the strength of the
hopping disorder $a$ and examine two quantities $\mathcal{F}$ and $C_{2}$.
It is shown that the system still possesses the $\eta $-pairing state even
though the small homogeneous $U$ and the disordered imaginary hopping
present, which is similar to that of the 1D system. Although the disorder $U$
affects the correlation of ground state, the correlator $C_{2}$ has a small
fluctuation around the value of uniform case supporting the
superconductivity of the ground state. Therefore, one can conclude that all
the results of 1D can be extended to 2D lattice system. It can be expected
that this conclusion is still valid for the higher dimensional bipartite
system.

\section{Transition from normal to $\protect\eta $-pairing ground states}

\label{transition} In this section, we focus on how does the ground state
transits from normal to superconductive state. To observe such a transition,
we consider a 1D Hubbard system with only two nearest neighbour (NN) sites
coupled through Hermitian hopping $t$. The corresponding Hamiltonian can be
given as
\begin{eqnarray}
H &=&-\sum_{j,\sigma =\uparrow ,\downarrow }t_{j}(c_{j,\sigma }^{\dagger
}c_{j+1,\sigma }+c_{j+1,\sigma }^{\dagger }c_{j,\sigma })  \notag \\
&&+U\sum_{j}n_{j,\uparrow }n_{j,\downarrow },  \label{h_t}
\end{eqnarray}%
where homogeneous $U$ is supposed and $t_{1}=t_{0}$ otherwise $t_{j}=0$. The
Hamiltonian still possesses the $\eta $ symmetry and hence supports the $%
\eta $-pairing eigenstate. However, such a state is not the ground state of
the system. Now we switch on the other coupling of the NN sites, which are
the imaginary hoppings rather than Hermitian hoppings, that is $t_{j}=it$
for $j\neq 1$. Fig. \ref{fig_gene} shows the variation of the low-lying
energy spectrum with respect to $it$. The $\eta $-pairing state is denoted
by the red line, which is suppressed to the ground state by the increase of
imaginary hopping. There exists a transition window in which the ground
state is transformed from a normal state to a superconducting state. We
perform the numerical simulation to demonstrate this process through the
correlator $C_{3}$. Evidently, $C_{3}$ undergoes a leap around the critical
point which leads to the divergence of the first order of derivative $%
\partial C_{3}/\partial t$. It witnesses the formation of the $\eta $%
-pairing ground state. Notice that all the conclusions can be extended to a
higher dimension. Before ending this section, we want to point out that the
imaginary hopping plays the key to achieve the superconducting ground state,
however, it does not mean that the system must have the $\eta $-pairing
ground state as long as the imaginary hopping is applied. The transition of
the ground state always requires a process such that there is a threshold
beyond which the system favors the superconductivity. Such property is
reminiscent of quantum phase transition, that is, the ground state will
experience a dramatic change when the system crosses the quantum phase
transition point. This findings paves the way to understand the $\eta $-spin
ferromagnetic state of the non-Hermitian strongly correlated system.

\section{Summary}

\label{summary} In summary, we have systematically studied the effect of the
non-Hermitian imaginary hopping on the low-lying energy spectrum of the
Hubbard model. The analytical solution within the two-particle subspace
shows that the introduction of the imaginary hopping results in a full
imaginary scattering spectrum and a flip of the bound pair spectrum
comparing to its Hermitian parent model. It indicates that the
particle-particle correlation suppresses the non-Hermiticity making the
ground state to be $\eta $-pairing state with ODLRO. The $\eta $ symmetry
plays the vital role in this mechanism. In the large negative $U$ limit, the
magnetism of the Hubbard model is altered fundamentally due to the interplay
between the particle-particle interaction and non-Hermitian imaginary
hopping. The ground state experiences a transition from normal to $\eta $%
-spin ferromagnetic states. Such a transition holds for any pair filled,
that is, the ground state in each invariant subspace is $\left( \eta
^{+}\right) ^{M}|\mathrm{Vac}\rangle $ with $M$ being the pairs of
particles. Through numerical simulation of 1D and 2D non-Hermitian Hubbard
system, we demonstrate that the $\eta $-pairing ground state can still
survival albeit a small negative $U$ presents. This evidence is robust
against disorder even if the system does not fulfill the $SO(4)$ symmetry.
Our results open a new avenue toward populating a $\eta $-pairing ground
state and suppressing antiferromagnetic correlation of $\eta $ spins in the
attractive Hubbard model.

\acknowledgments We acknowledge the support of the National Natural Science
Foundation of China (Grants No. 11975166, and No. 11874225). X.Z.Z. is also
supported by the Program for Innovative Research in University of Tianjin
(Grant No. TD13-5077).

\appendix
\label{appendix}

\section{Two-particle solutions}

In this section, we show the detailed caculation for the two-particle
solution in each invariant subspace. For the simplicity, we only focus on
the solutions in subspaces $\left( 0,\text{ }0\right) $ and $\left( 1,\text{
}0\right) $, since the solution in subspace $\left( 1,\text{ }\pm 1\right) $
can be obtained directly from that in subspace $\left( 1,\text{ }0\right) $
by operator $s^{\pm }$. A two-particle state can be given as%
\begin{equation}
\left\vert \varphi _{K}^{\pm }\right\rangle =\sum_{r}f_{K,k}^{\pm }\left(
r\right) \left\vert \phi _{r}^{\pm }\left( K\right) \right\rangle \text{, }%
\left( f_{K}^{+}\left( 0\right) =f_{K,k}^{-}\left( -1\right) =0\right) ,
\end{equation}%
where $r$ denotes the relative distance between the two particles and the
wave function $f_{K,k}^{\pm }\left( r\right) $\ obeys the Schr\"{o}dinger
equations%
\begin{eqnarray}
Q_{r}^{K}f_{K,k}^{+}\left( r+1\right) +Q_{r-1}^{K}f_{K,k}^{+}\left(
r-1\right) + &&  \notag \\
\lbrack \left( -1\right) ^{n}Q_{r}^{K}\delta _{r,N_{0}}-\varepsilon
_{K}]f_{K,k}^{+}\left( r\right) =0, &&  \label{S+}
\end{eqnarray}%
and%
\begin{eqnarray}
Q_{r}^{K}f_{K,k}^{-}\left( r+1\right) +Q_{r-1}^{K}f_{K,k}^{-}\left(
r-1\right) + &&  \notag \\
\lbrack U\delta _{r,0}+\left( -1\right) ^{n}Q_{r}^{K}\delta
_{r,N_{0}}-\varepsilon _{K}]f_{K,k}^{-}\left( r\right) =0, &&  \label{S-}
\end{eqnarray}%
with $N_{0}=\left( N-1\right) /2$ and the eigen energy $\varepsilon _{K}$\
in the invariant subspace indexed by $K$. Here factor $Q_{r}^{K}=-2\sqrt{2}%
it\cos \left( K/2\right) $ for $r=0$ and $-2it\cos \left( K/2\right) $ for $%
r\neq 0$, respectively. $U$ appears in the $(0,0)$ subspace and therefore
admits the bound pair solution. In the large $N$ limit, we can neglect the
effect of on-site potential $\left( -1\right) ^{n+1}2it\cos \left(
K/2\right) $ at $N_{0}$th site. The solution of (\ref{S-}) is equivalent to
that of the single-particle semi-infinite tight-binding chain system with
nearest-neighbour (NN) hopping amplitude $Q_{j}^{K}$, and on-site potentials
$U$ at $0$th site, respectively. Moreover the solution of (\ref{S+})
corresponds to the same chain with infinite $U$. In this scenario, the bound
state solution $\left\vert \varphi _{K}^{\mathrm{b}}\right\rangle
=\sum_{r}f_{K}^{-}\left( r\right) \left\vert \phi _{r}^{-}\left( K\right)
\right\rangle $ can be determined by substituting the ansatz
\begin{equation}
f_{K}^{-}\left( j\right) =\left\{
\begin{array}{c}
1/\sqrt{2}\text{, }j=0 \\
e^{-\beta j}\text{, }j\neq 0%
\end{array}%
\right.
\end{equation}%
into the following equivalent Hamiltonian
\begin{equation}
H_{\text{eq}}^{K}=U\left\vert 0\right\rangle \left\langle 0\right\vert
+\sum_{i=0}^{\infty }\left( Q_{i}^{K}\left\vert i\right\rangle \left\langle
i+1\right\vert +\text{H.c.}\right) .
\end{equation}%
Straightforward algebra shows that $\beta =\ln [(-U\pm \sqrt{U^{2}+4\lambda
_{K}^{2}})/2\lambda _{K}]$ where $\lambda _{K}=2it\cos \left( K/2\right) $
and $\pm $ denotes negative and positive $U$, respectively. Correspondingly,
the energy of the bound pair is
\begin{equation}
\epsilon _{K}=\mathrm{sgn}\left( U\right) \sqrt{U^{2}-16t^{2}\cos ^{2}\left(
K/2\right) }.
\end{equation}%
For the case of negative $U$, the lowest energy of bound pair is $\epsilon
_{\pi }=-U$ locating on the subspace with $K=\pi $. As such the
corresponding eigenstate is $\left\vert \phi _{0}^{-}\left( K\right)
\right\rangle $ that represents a $\eta $-pairing state in the coordinate
space with the form of $\left( \eta ^{+}\right) /\sqrt{N}|\mathrm{Vac}%
\rangle $.

\section{Simple example of two-site case for the effective Hamiltonian $H_{%
\mathrm{eff}}$}

In this subsection, we provide a detailed calculation of the two-site case
for the effective Hamiltonian $H_{\mathrm{eff}}$ which may shed light to
obtain the effective Hamiltonian (\ref{heff}). In the simplest two-site
case, $P_{0}=\sum_{\alpha \in \text{\textrm{d.o.}}}|\alpha \rangle \langle
\alpha |$ is the projection operator to the doublon subspace spanned by the
configuration $\left\{ |\text{\textrm{x}}0\rangle ,|0\text{\textrm{x}}%
\rangle \right\} $, and $P_{1}=1-P_{0}=\sum_{a\notin \text{\textrm{d.o.}}%
}|a\rangle \langle a|$ is the complementary projection. Here the
abbreviation \textrm{d.o.} means the doubly occupied subspace and $|$\textrm{%
x}$0\rangle =c_{1,\uparrow }^{\dagger }c_{1,\downarrow }^{\dagger }|$\textrm{%
Vac}$\rangle $, $|0$\textrm{x}$\rangle =c_{2,\uparrow }^{\dagger
}c_{2,\downarrow }^{\dagger }|$\textrm{Vac}$\rangle $. The first term of Eq.
(\ref{heff}) clear gives $P_{0}H_{0}P_{0}=U$. The second term can be
simplified by noting: (i) the unperturbed energy $E_{0}$ is $U$; (ii) $%
P_{1}H^{\prime }P_{0}$ annihilates the doubly occupied site. Then $H_{%
\mathrm{eff}}$ can be written as
\begin{eqnarray}
H_{\mathrm{eff}} &=&U+\sum_{\alpha ,\beta \in \text{d.o.}}\sum_{a,b\notin
\text{d.o.}}|\alpha \rangle \langle \alpha |H^{\prime }|a\rangle \langle a|
\notag \\
&&\times \frac{1}{U-H_{0}}|b\rangle \langle b|H^{\prime }|\beta \rangle
\langle \beta |  \notag \\
&=&U+\frac{1}{U}\sum_{\alpha ,\beta \in \text{d.o.}}\langle \alpha |\left(
H^{\prime }\right) ^{2}|\beta \rangle |\alpha \rangle \langle \beta |.
\end{eqnarray}%
The second term describes the virtual exchange of the fermions. The
non-Hermitian imaginary hopping brings about an additional sign to this
process yielding that
\begin{equation}
H_{\mathrm{eff}}=U-\frac{2t^{2}}{U}\left( |\text{\textrm{x}}0\rangle \langle
0\text{\textrm{x}}|+|0\text{\textrm{x}}\rangle \langle \text{\textrm{x}}0|+|%
\text{\textrm{x}}0\rangle \langle \text{\textrm{x}}0|+|0\text{\textrm{x}}%
\rangle \langle 0\text{\textrm{x}}|\right) .
\end{equation}%
Combining the cases in the subspaces of $|\mathrm{xx}\rangle $ and $|$%
\textrm{Vac}$\rangle $, the pseudo spin Hamiltonian can be given by the
non-Hermitian Heisenberg-like model
\begin{equation}
H_{\mathrm{eff}}=MU+\frac{4t^{2}}{U}\left( \bm{\eta }_{1}\cdot \bm{\eta }%
_{2}-\frac{1}{4}\right) ,
\end{equation}%
where $M$ can be $0$, $1$, and $2$ denoting the number of pairs of the
doublon subspace. Evidently, the ground state of $H_{\mathrm{eff}}$ is the $%
\eta $-spin ferromagnetic state with the form of $\left( \eta ^{+}\right)
^{2}|$\textrm{Vac}$\rangle $.


\end{document}